# Resonance Laser Ionization Spectroscopy of Selenium

R. Li[1*], Y. Liu[2], M. Mostamand[1,3], and J. Lassen[1,3,4]

[1]TRIUMF - Canada's Particle Accelerator Centre, Vancouver, BC, V6T 2A3, Canada
[2]Facility for Rare Isotope Beams, Michigan State University, East Lansing, MI 48824, USA
[3]Department of Physics and Astronomy, University of Manitoba, Winnipeg, MB, R3T 2N2, Canada
[4]Department of Physics, Simon Fraser University, Burnaby, BC, V5A 1S6, Canada

**Abstract**

Photoionization spectra of Se have been studied by step-wise resonance laser ionization. The Rydberg series $4s^24p^3(^4S)np\ ^3P_{0,1,2}$ and $4s^24p^3(^4S)np\ ^5P_{1,2,3}$ were measured via different excitation schemes. Using the Rydberg series $4s^24p^3(^4S)np\ ^3P_2$ with $n$=15-33, the ionization potential of Se was determined with improved precision to 76658.15(2)$_{stat}$(4)$_{sys}$ cm$^{-1}$, which resolved the discrepancy in previous literatures. Autoionizing (AI) spectra between the IP and two neighboring converging limits of the Se ionic states $4s^24p^3(^2D_{3/2})$ and $4s^24p^3(^2D_{5/2})$ were obtained. In total eight AI Rydberg series have been observed, measured and assigned.

## 1. Introduction

Selenium (Se) is a chalcogen element that shows atomic properties similar to its homologues sulfur (S) and tellurium (Te). With four valence electrons in an open $p$ shell rich resonance structures with strong configuration interactions can be observed. With increasing atomic mass relativistic effects also increase and the spin-orbit interaction starts to dominate. Considering multiple types of interactions and several valence electrons, the theoretical description of Se and Te is challenging [1]. Moreover, the experimental spectroscopic information on Se is rather limited [2], due to the difficulty to produce an atomic vapor and to access the UV and deep UV resonances of Se. The latest reported spectroscopic data were measured in the 1980s with traditional emission and absorption spectroscopy. The lack of experimental data further hindered the theoretical study and atomic modeling of these elements. Meanwhile in astrophysics studies, Se is expected to play an important role in the transition from light to heavy elements in neutron capture nucleosynthesis. It was searched for but had never been observed in the stellar spectra including the Sun until recently [3]. With the imaging spectrograph onboard the Hubble space telescope, selenium was observed in an ancient halo star [4]. Therefore improving and extending the atomic spectroscopy data of Se is needed and desired by the astrophysics community.

Starting in the 1920s, early studies on Se atomic structures were implemented via arc spectroscopy [5]. Ruedy and Gibbs [6,7] observed the emission spectrum of Se from 130-1040 nm and classified 510 lines and 130 levels. However only 42 levels of their results agreed with Meissner's work [8], which reported 111 levels at nearly the same time. To clarify the discrepancy, the spectroscopic data of Se was later revised and extended in the 1970-80s: Erikson precisely measured the ground state fine structures [9]; Morillon and Vergès[10] thoroughly measured the Se I spectrum in the range 884 - 2513 nm using a Fourier transform spectrometer, with a level energy precision of ±0.003cm$^{-1}$. They also used the $4p^3(^4S)nf\ ^5F_5$ series (n=4-10) to determine the ionization potential (IP) of Se to be 78658.12(10) cm$^{-1}$.

---

* corresponding email: ruohong@triumf.ca

In parallel, absorption spectroscopy of Se started to thrive, since the technique of flash photolysis [11] solved the problem of generating atomic Se vapors. When using traditional thermal vaporization, Se has a strong tendency to form molecules, such as $Se_2$, $Se_4$ and $Se_8$. In the photolysis technique Se is vaporized by high voltage discharges through a flash lamp, which significantly dissociates Se molecules into atoms. Lindgren and Palenius [12] first used this method on Se and reported 34 new lines in wavelength range 156-199.5 nm. They revised the Morillon and Vergès's results and claimed a systematic energy shift of $0.23 \pm 0.02$ cm$^{-1}$. Consquently they revised the IP value to 78658.35(12) cm$^{-1}$. Cantú and Mazzoni [13] extended the study to the vacuum ultraviolet region and measured absorption spectra in the VUV range of 80-230 nm. Odd-parity Rydberg series *$4p^3(^4S)ns$ $^5S_2$, $4p^3(^4S)ns$ $^3S_1$, $4p^3(^4S)nd$ $^3D_{1,2,3}$* for *n*=4-29 were measured. They also conducted the first study of odd-parity autoionizing (AI) states between the IP and the limit *$4p^3(^2D_{3/2})$* [14]. To eliminate the possibility of molecular impurities in Se vapor, Gibson et al. [15] used a chemical reaction of H and $H_2Se$ to generate atomic Se and studied the photoionization spectra of odd-parity AI states above the IP and below the *$4p^3(^2P_{3/2})$* limit.

In this work, step-wise laser resonant ionization spectroscopy is used to study even-parity Rydberg states and odd-parity AI states. Selective ionization using multiple resonant excitation steps eliminates false signals from molecular impurities, and significantly reduces the ambiguity in assigning level configurations. With cleaner spectra and a higher signal to noise ratio, more details of Rydberg spectra can be observed especially for high-*n* Rydberg states, which gives a better assessment for determining the IP.

## 2. Experimental Setup

The experiment was performed at TRIUMF's offline laser ion source test stand (LIS-STAND) [16] shown in Fig. 1. Beads (shots) of elemental Se (~5 mg each) were ground into small pieces and loaded into a Ta crucible. The crucible is a 20 mm long Ta tube with 3 mm inner diameter that can be resistively heated. With a current of up to 300 A it is possible to heat the crucible to temperatures above 2000 °C. Se has a strong propensity to form molecules or dimers rather than atoms, similar to its lighter and heavier homologues O and Te. To efficiently break these molecular bonds a crucible temperature above 1600 °C was needed, as determined from experimental observations. The Se atoms in the crucible interacted with multiple laser beams that stepwise excited the valence electron of the atom into higher electronic levels and eventually ionizing the atom. The resulting ions were guided by electric fields through a 50 mm long radio frequency quadrupole (RFQ) ion guide [17], then extracted and accelerated to 10 keV. The fast ion beam was focused by an Einzel lens and then deflected 90° vertically, which geometrically allowed the laser beams to access the ionization region inside the crucible. In the ~2 m vertical section, the ion beam was refocused and deaccelerated to below 50 eV beam energy before entering a quadrupole mass spectrometer (QMS, ABB Extrel MAX-300) operated at unit mass resolution or better ($\Delta$m <1amu). The mass filtered ions were collected on a conversion dynode biased at -5 kV, where secondary electrons were generated and detected downstream by a channel electron multiplier biased at -2 kV.

In the neighboring laser lab, a 10 kHz repetition rate, frequency doubled Nd:YAG laser with 28 W output power is used to pump three titanium:sapphire (Ti:Sa) lasers simultaneously. The output power of the Ti:Sa lasers is typically 1-2 W with a 680-980 nm tunable range. The pulse length of the Ti:Sa lasers is ~50 ns (FWFH). Two Ti:Sa lasers are wavelength tuned via birefringent filters (BRF), and the other is grating-tuned. The BRF-tuned lasers provide excellent stability and high output power. Therefore they are typically used to excite known transitions, such as the first/second excitation steps. The grating-tuned laser is utilized for laser spectroscopic investigations to search for unknown atomic structures. An outstanding feature of the grating-tuned Ti:Sa laser is its capability to scan continuously across a broad wavelength range 700-930 nm [18]. This significantly simplifies laser scan procedures and boosts the search efficiency. The wavelength range of the laser system can been extended to the blue and UV spectral range by adding higher harmonic frequency generation stages. For the Se excitation scheme, as shown in Fig. 2, the UV first step excitation was achieved by quadrupling the light from a BRF-tuned Ti:Sa laser. The laser was first intra-cavity doubled by inserting a BBO crystal inside the laser cavity [19] to provide 400 mW blue laser light with a near Gaussian beam profile. This light was then frequency doubled externally in another BBO crystal. This resulted ~20 mW UV light for two Se atomic transitions at 207.546 nm and 206.345 nm (all wavelengths stated in this paper are measured in vacuum) as the first excitation steps from the ground state 0.00 cm$^{-1}$ or the metastable state 2534.36 cm$^{-1}$, respectively. To search for efficient second/third excitation steps, both fundamental (infrared) and second harmonic (blue) wavelength ranges of the grating tuned Ti:Sa laser were scanned. The blue laser scanning was obtained by an automated external frequency doubling system. A type-I cut LBO crystal was mounted on a motorized rotary stage with the fundamental output of the grating-tuned laser being focused inside the crystal by a plano-convex lens of $f$=50 mm. The phase matching condition during laser wavelength scans was maintained by automatically rotating the crystal optical axis relative to the incident laser beam. However, the walk-off of the laser beam passing through the 8 mm long crystal was considerable at wavelengths far off the crystal cut wavelength. In order to maintain the spatial overlap of the scanning laser with other lasers inside the 3-mm diameter crucible 5 m away from the laser table, a commercial beam stabilization system (TEM Messtechnik BeamLock-4D) was employed. The system includes two automated mirror mounts and two position sensitive detectors, which gives a satisfactory precision to automatically correct any beam pointing drifts during long laser wavelength scans. The details of this system have been described in previous work [20].

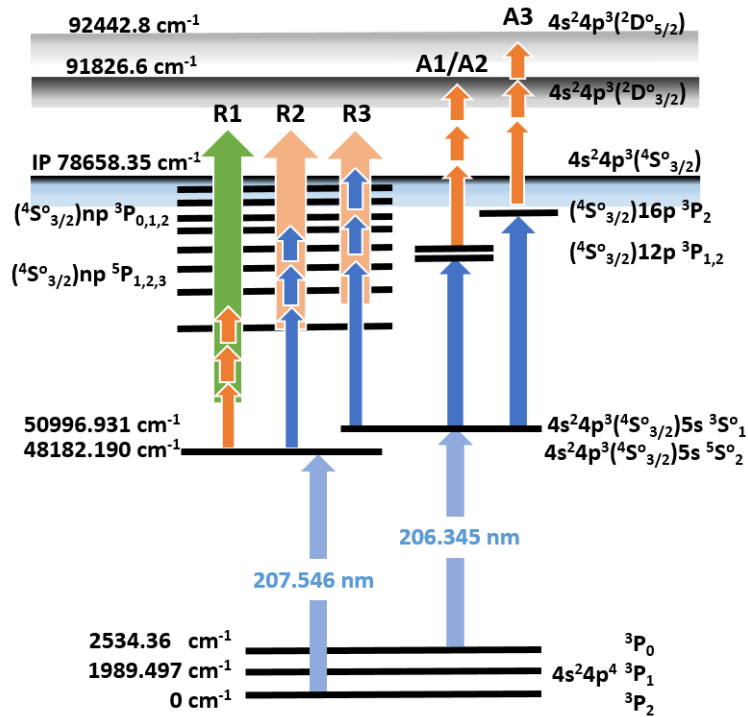

**Fig. 1.** Laser ionization schemes of atomic selenium investigated in the present work. The level energies and IP are taken from NIST database and presented to their best known precisions [2].

To efficiently ionize Se atoms, laser pulses for all excitation/ionization steps have to overlap in the ionization region spatially and temporally. The laser pulses can be synchronized by adjusting the pump laser power and/or by using Pockels cells inside the laser cavity. The spatial superposition of laser beams was achieved by utilizing polarizing beam splitters and dichroic mirrors. All the laser beams were individually collimated and expanded 2-4 times in order to be efficiently focused into the ionization region via a $f$=5 m uncoated 50.8 mm diameter lens. The laser wavelengths were measured and monitored by a commercial wavemeter (HighFinesse WS/6) multiplexed with a 4-channel MEMS opto-mechanical switcher. To ensure measurement accuracy, the wavemeter was routinely calibrated to a polarization stabilized HeNe laser with a wavelength accuracy of $10^{-8}$ (Melles Griot 05 STP 901/903, identical to the Spectra-Physics model 117A). The systematic uncertainty is 0.02 cm$^{-1}$ with 3-σ criterion.

## 3. Experimental procedures and results

All excitation schemes investigated in this work are shown in Fig. 1, where schemes R1- R3 aimed to investigate Rydberg states, and schemes A1-A3 were intended to investigate AI states.

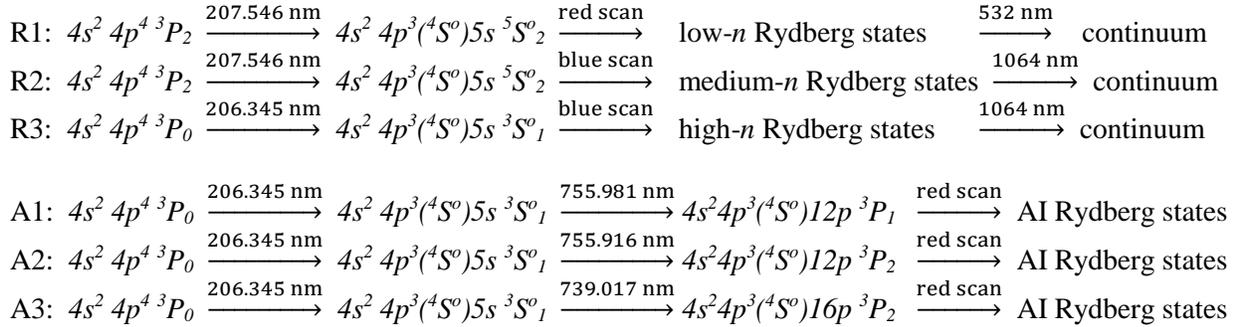

### 3.1 Rydberg spectra via scheme R1/R2/R3

Excitation schemes R1, R2 and R3 allow to access the different energy ranges of Rydberg states below the IP and to investigate the even-parity Rydberg series $4s^2 4p^3(^4S)np\ ^5P_{1,2,3}$ and $4s^2 4p^3(^4S)np\ ^3P_{0,1,2}$. Schemes R1 and R2 excited the Se atoms initially in the ground state, and scheme R3 excited from the metastable state at 2534.36 cm$^{-1}$. The Boltzmann population distribution at the experimental crucible temperature of 1800 °C is ~70% for the ground state and ~12% for the metastable state. The purpose to excite Se atoms from the metastable state was to access high-$n$ Rydberg states, although the ion signal would be reduced. This could be a problem as the photoexcitation rate to Rydberg states decreases with increasing principal number $n$, scaling with $n^{-3}$. In this experiment a sufficient signal to noise ratio was achieved up to Rydberg states with $n=36$. The first excitation steps of scheme R1/R2 (207.504 nm) and R3 (206.345 nm) excited Se atoms to the intermediate $(^4S)5s\ ^5S_2$ state at 48182.190 cm$^{-1}$ and $(^4S)5s\ ^3S_1$ state at 50996.931 cm$^{-1}$ respectively. The second steps of scheme R1/R2/R3 were provided by either fundamental or frequency doubled light of the grating-tuned laser. The typical power for fundamental laser (red scan) was 0.8-2 W, and for frequency doubled laser (blue scan) was 50-300 mW dependent on the wavelength. Fig. 2 shows the photoionization spectra obtained via scheme R3 for the energy range 78050-78600 cm$^{-1}$, presented as ion signal versus level energy. The measured level energies, which are presented in Tab.1 and Tab.2, were extracted by fitting the resonance peaks using a Gaussian profile. Each spectrum was measured more than once to check the reproducibility and the statistic uncertainty of the resonance centroids. Therefore the level energies presented in Tab.1 and Tab.2 are the averages of 2-3 measurements. The uncertainty of the level energy is 0.15 cm$^{-1}$, including the measurement statistics and the systematic uncertainty of the experimental system [21,22].

The measured level energies are also presented in a Lu-Fano plot (Fig. 3). The effective quantum number $n^*$ and quantum defect $\delta$ are both calculated from Rydberg-Ritz formula, which describes how a Rydberg series $E_n$ converges to an ionization limit $E_{limit}$:

$$E_n = E_{limit} - \frac{R_M}{n^{*2}} = E_{limit} - \frac{R_M}{[n-\delta(n)]^2} \tag{1}$$

$$\delta(n) = \delta_0 + \frac{a}{(n-\delta_0)^2} + \frac{b}{(n-\delta_0)^4} \tag{2}$$

Here $R_M$ is the mass-corrected Rydberg constant 109736.56 cm$^{-1}$ for $^{80}$Se, the most abundant stable Se isotope (49.6%) mass-selected to measure the Rydberg spectra. For the Rydberg series observed by scheme R1, R2 and R3, $E_{limit}$ corresponds to the ionization potential (IP). Due to core penetration and/or polarization effects, the quantum defect is not constant for low-$n$ Rydberg members. Therefore, the higher orders of the Ritz expansion (with coefficients $a$ and $b$) are needed in the case that an obvious $\delta$ deviation is present. For instance, in Fig. 3 the $\delta$ deviation of the series $np\ ^3P_{1,2}$ starts at $n=15$, and increase towards lower-n. Around $n=35$, the series $np\ ^3P_2$ is perturbed, showing a local $\delta$ deviation from $\delta_0 \sim 2.45$ (mod$_1[\delta]$=0.45 in Fig. 3) and the rapid resonance signal drop (Fig. 2) after $n=32$. This causes no more resonances observed for $n>36$. Without perturbations, up to $n=54$ of the corresponding Rydberg states had been observed for the homologue element Te [20].

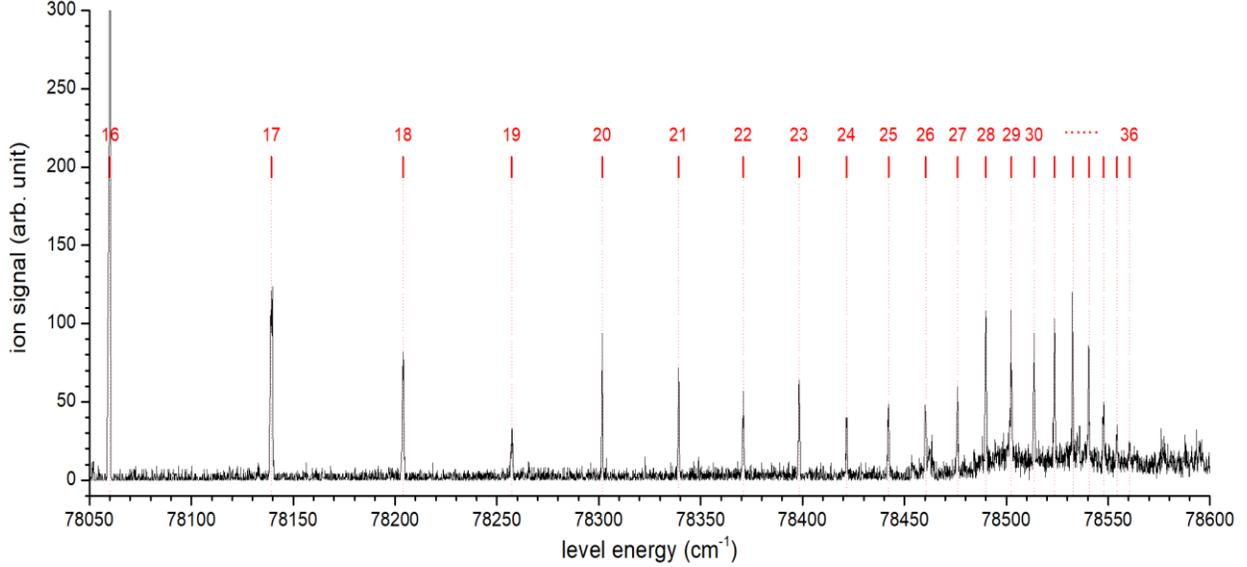

**Fig. 2.** Photoionization spectra of the Rydberg series $4s^24p^3(^4S)np\ ^3P_2$ observed via scheme R3. The principal quantum number $n$ is marked with drip lines on the resonances. The background rise at 78500 cm$^{-1}$ was double-checked and proved to be non-reproducible, hinting at an uneven vaporization of the sample.

Using the Rydberg series $4s^24p^3(^4S)np\ ^3P_2$, the IP value can be extracted via Eq. (1) and (2). Fig. 4 shows the fitted curve with the experimental data points and the residual of the nonlinear least squares fitting. The Rydberg members with $n>14$ were chosen for the fitting to avoid the significant core effect in the low $n$ range. During the fitting, different orders of the Ritz expansion eq. (2) were checked. All of them gave a consistent value of 78658.15(2) cm$^{-1}$ with an even residual distribution < 0.15 cm$^{-1}$. As stated in the introduction, the currently accepted IP value of 78658.35(12) cm$^{-1}$ [2, 12] was revised by Lindgren and Palenius from the orignal result of 78658.12(10) cm$^{-1}$ measured by Morillon and Vergès[10]. This revision was based on a systematic shift of 0.23±0.02 cm$^{-1}$ of their measurements from that of Morillon and Vergès. It should be noted that NIST accepted the former's IP value but the latter's level energies. To resolve the discrepancy, We measured the level energy of $4s^2\ 4p^3(^4S^o)5s\ ^3S^o_1$ several times and obtained 50996.89(15) cm$^{-1}$. It agrees better with Morillon's original 50996.931(3) cm$^{-1}$ instead of Lindgren's value of 50997.15 cm$^{-1}$. This level was also the intermediate level we used to scan Rydberg series. Therefore it is critical to know its accurate level energy in calculating the energies of Rydberg states. In our anaylisis we used Morillon's value 50996.931(3) cm$^{-1}$, which is also the present NIST-accepted value. Considering the uncertainty of this level and the systematic uncertainty 0.04 cm$^{-1}$ from the wavemeter [23], the resulting IP value as determined from our data is 78658.15(2)$_{stat}$(4)$_{sys}$ cm$^{-1}$, which agrees with Morillon's value.

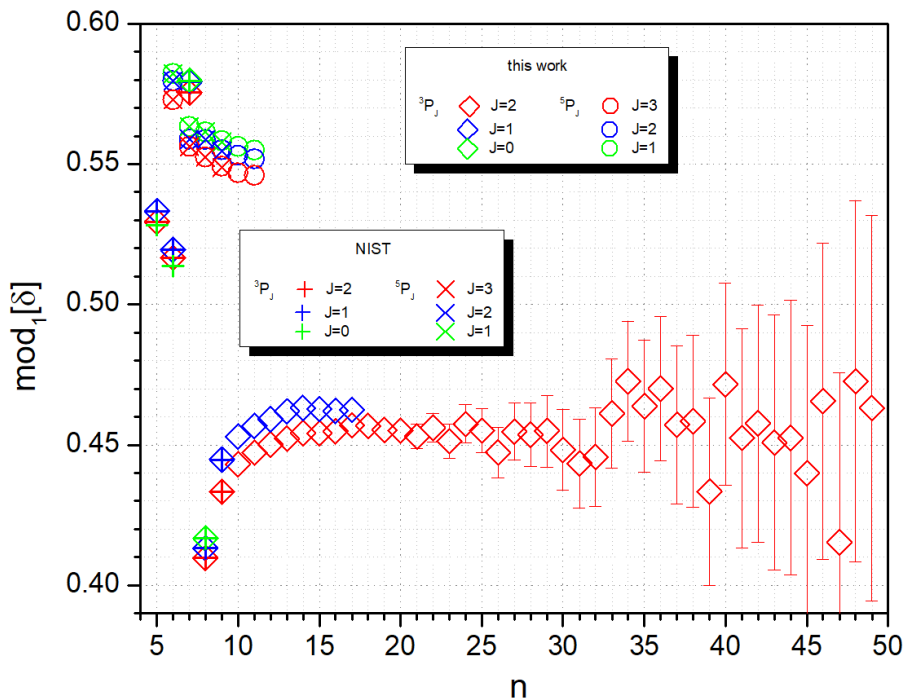

**Fig. 3.** Lu-Fano plot for the observed even-parity $4s^24p^3(^4S)np$ Rydberg series in this work comparing with the corresponding NIST data [2]. The y axis presents $\mathrm{mod}_1[\delta]$, *e.g.*, the decimal part of the quantum defect $\delta$, which is calculated with the IP value 78658.15 cm$^{-1}$ determined in this work. For all $np$ series, $\delta = 2 + \mathrm{mod}_1[\delta]$. The fine structure of the triplets $^3P_{0,1,2}$ and $^5P_{1,2,3}$ are marked in different colors, as same as the NIST data. For clarity error bars smaller than the symbol size are hidden.

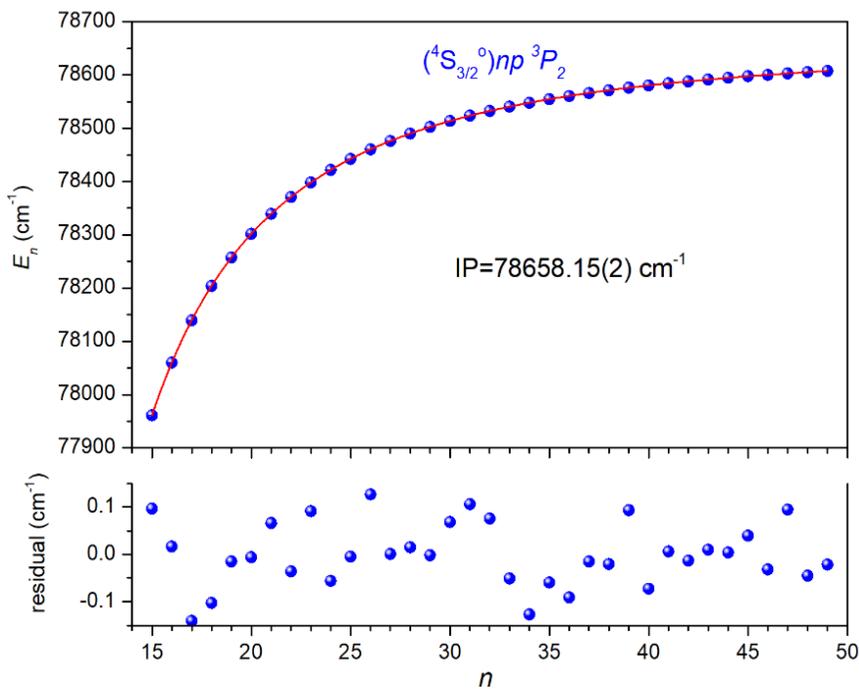

**Fig. 4.** IP determination by Rydberg series $4s^24p^3(^4S)np\ ^3P_2$ using the data $n$=15-33.

**Table 1.** Series $4s^24p^3(^4S)np\ ^3P_{0,1,2}$ converging to the IP. The level energies of this work were measured via schemes R1, R2 and R3. The previously known levels from NIST database [2] are also presented for comparison. The quantum defect $\delta$ is calculated with the IP value 78658.15 cm$^{-1}$ determined in this work.

| | J=2 | | | J=1 | | | J=0 | | |
|---|---|---|---|---|---|---|---|---|---|
| | this work | | NIST | this work | | NIST | this work | | NIST |
| n | $E_n$ (cm$^{-1}$) | $\delta$ | $E_n$ (cm$^{-1}$) | $E_n$ (cm$^{-1}$) | $\delta$ | $E_n$ (cm$^{-1}$) | $E_n$ (cm$^{-1}$) | $\delta$ | $E_n$ (cm$^{-1}$) |
| 5 | 60677.49 | 2.53 | 60677.388 | 60622.41 | 2.53 | 60622.302 | | | 60695.989* |
| 6 | 69614.09 | 2.52 | 69614.114 | 69599.48 | 2.52 | 69599.464 | | | 69629.529 |
| 7 | 73052.43 | 2.58 | 73052.408 | 73042.92 | 2.58 | 73042.922 | 73041.95 | 2.58 | 73041.948 |
| 8 | 75146.64 | 2.41 | 75146.71 | 75142.24 | 2.41 | 75142.32 | 75137.76 | 2.42 | 75137.89 |
| 9 | 76113.19 | 2.43 | 76113.36 | 76104.30 | 2.45 | 76104.53 | | | |
| 10 | 76736.50 | 2.44 | | 76731.46 | 2.45 | | | | |
| 11 | 77157.93 | 2.45 | | 77154.59 | 2.46 | | | | |
| 12 | 77454.89 | 2.45 | | 77452.62 | 2.46 | | | | |
| 13 | 77671.79 | 2.45 | | 77669.97 | 2.46 | | | | |
| 14 | 77834.96 | 2.45 | | 77833.67 | 2.46 | | | | |
| 15 | 77960.95 | 2.45 | | 77960.00 | 2.46 | | | | |
| 16 | 78060.03 | 2.45 | | 78059.39 | 2.46 | | | | |
| 17 | 78139.29 | 2.46 | | 78138.90 | 2.46 | | | | |
| 18 | 78203.92 | 2.46 | | | | | | | |
| 19 | 78257.25 | 2.46 | | | | | | | |
| 20 | 78301.65 | 2.46 | | | | | | | |
| 21 | 78339.14 | 2.45 | | | | | | | |
| 22 | 78370.85 | 2.46 | | | | | | | |
| 23 | 78398.26 | 2.45 | | | | | | | |
| 24 | 78421.69 | 2.46 | | | | | | | |
| 25 | 78442.25 | 2.46 | | | | | | | |
| 26 | 78460.33 | 2.45 | | | | | | | |
| 27 | 78476.00 | 2.45 | | | | | | | |
| 28 | 78490.00 | 2.45 | | | | | | | |
| 29 | 78502.42 | 2.45 | | | | | | | |
| 30 | 78513.59 | 2.45 | | | | | | | |
| 31 | 78523.58 | 2.44 | | | | | | | |
| 32 | 78532.51 | 2.45 | | | | | | | |
| 33 | 78540.48 | 2.46 | | | | | | | |
| 34 | 78547.75 | 2.47 | | | | | | | |
| 35 | 78554.49 | 2.46 | | | | | | | |
| 36 | 78560.54 | 2.47 | | | | | | | |
| 37 | 78566.18 | 2.46 | | | | | | | |
| 38 | 78571.28 | 2.46 | | | | | | | |
| 39 | 78576.08 | 2.43 | | | | | | | |
| 40 | 78580.23 | 2.47 | | | | | | | |
| 41 | 78584.30 | 2.45 | | | | | | | |
| 42 | 78587.97 | 2.46 | | | | | | | |
| 43 | 78591.41 | 2.45 | | | | | | | |
| 44 | 78594.58 | 2.45 | | | | | | | |
| 45 | 78597.57 | 2.44 | | | | | | | |
| 46 | 78600.25 | 2.47 | | | | | | | |
| 47 | 78602.94 | 2.42 | | | | | | | |
| 48 | 78605.21 | 2.47 | | | | | | | |
| 49 | 78607.48 | 2.46 | | | | | | | |

Note: * corrected value from [10]

**Table 2.** Series $4s^24p^3(^4S)np\ ^5P_{1,2,3}$ converging to the IP. These levels were observed via schemes R1 and R2. The energies measured in this work are compared to the previous data from NIST database [2]. The quantum defect $\delta$ is calculated with the IP value 78658.15 cm$^{-1}$ determined in this work.

| | J=3 | | | J=2 | | | J=1 | | |
|---|---|---|---|---|---|---|---|---|---|
| | this work | | NIST | this work | | NIST | this work | | NIST |
| n | $E_n$ (cm$^{-1}$) | $\delta$ | $E_n$ (cm$^{-1}$) | $E_n$ (cm$^{-1}$) | $\delta$ | $E_n$ (cm$^{-1}$) | $E_n$ (cm$^{-1}$) | $\delta$ | $E_n$ (cm$^{-1}$) |
| 6 | 69314.41 | 2.57 | 69314.405 | 69277.59 | 2.58 | 69277.616 | 69263.25 | 2.58 | 69263.236 |
| 7 | 73101.06 | 2.57 | 73101.107 | 73094.33 | 2.56 | 73094.379 | 73083.09 | 2.56 | 73083.111 |
| 8 | 74960.23 | 2.57 | 74960.219 | 74951.80 | 2.56 | 74951.85 | 74948.10 | 2.56 | 74948.08 |
| 9 | 76021.04 | 2.57 | 76021.52 | 76016.26 | 2.56 | 76016.64 | 76013.68 | 2.56 | 76013.93 |
| 10 | 76682.73 | 2.57 | | 76679.43 | 2.55 | | 76677.74 | 2.56 | |
| 11 | 77122.84 | 2.57 | | 77120.69 | 2.55 | | 77119.53 | 2.56 | |

### *3.2 AI state spectra via scheme A1, A2 and A3*

In previous studies of the AI states in chalcogen elements between the core ground state $(^4S^o_{3/2})$ and the second excited state $(^2D^o_{5/2})$, oxygen is the best studied and exhibits only sharp resonances due to radiative emission competition [24]. Starting from sulfur, AI photoionization resonances started to show distinct narrow and broad figures, depending on how strongly they couple to the continuum. With increasing mass the growing spin-orbit interaction breaks down *LS* selection rules. This increases the complexity in AI photoionization spectra for the experimental spectra interpretation as well as theoretical atomic modeling. Chen et al. used *LS*-coupled eigenchannel R-matrix combined with the *jj-LS* frame transformation to calculate the AI photoionization spectrum of S, which agreed well with Gibson's experimental data [25]. However the same method gave only global but not detailed agreement with measurements on Se and the agreement turned remarkably worse for Te, which implies the limitation of *jj-LS* frame transformation as a method including the spin-orbit effects [26]. Our recent experimental result of Te AI states using laser ionization spectroscopy showed a good description of observed states with *jj*-coupling [20,27]. This motivated us to investigate Se photoionization AI spectra in this energy range to bridge the configuration interaction transition, and to reveal systematic trends [24].

To access the AI Rydberg states converging to the limit 91826.6 cm$^{-1}$ of the core state $4s^24p^3(^2D^o_{3/2})$, and the limit 92442.6 cm$^{-1}$ of the core state $4s^24p^3(^2D^o_{5/2})$, three resonance steps are needed for the valence electron to obtain enough energy. In addition the 206.345 nm transition was used as the first excitation step, at the expense of the initial state population. The photoionization spectra obtained from schemes A1-A3 are shown in Fig. 5, excited from the intermediate states $4s^24p^3(^4S^o)12p\ ^3P_1$ at 77452.62 cm$^{-1}$, $4s^24p^3(^4S^o)12p\ ^3P_2$ at 77454.89 cm$^{-1}$ and $4s^24p^3(^4S^o)16p\ ^3P_2$ at 78060.03 cm$^{-1}$, respectively.

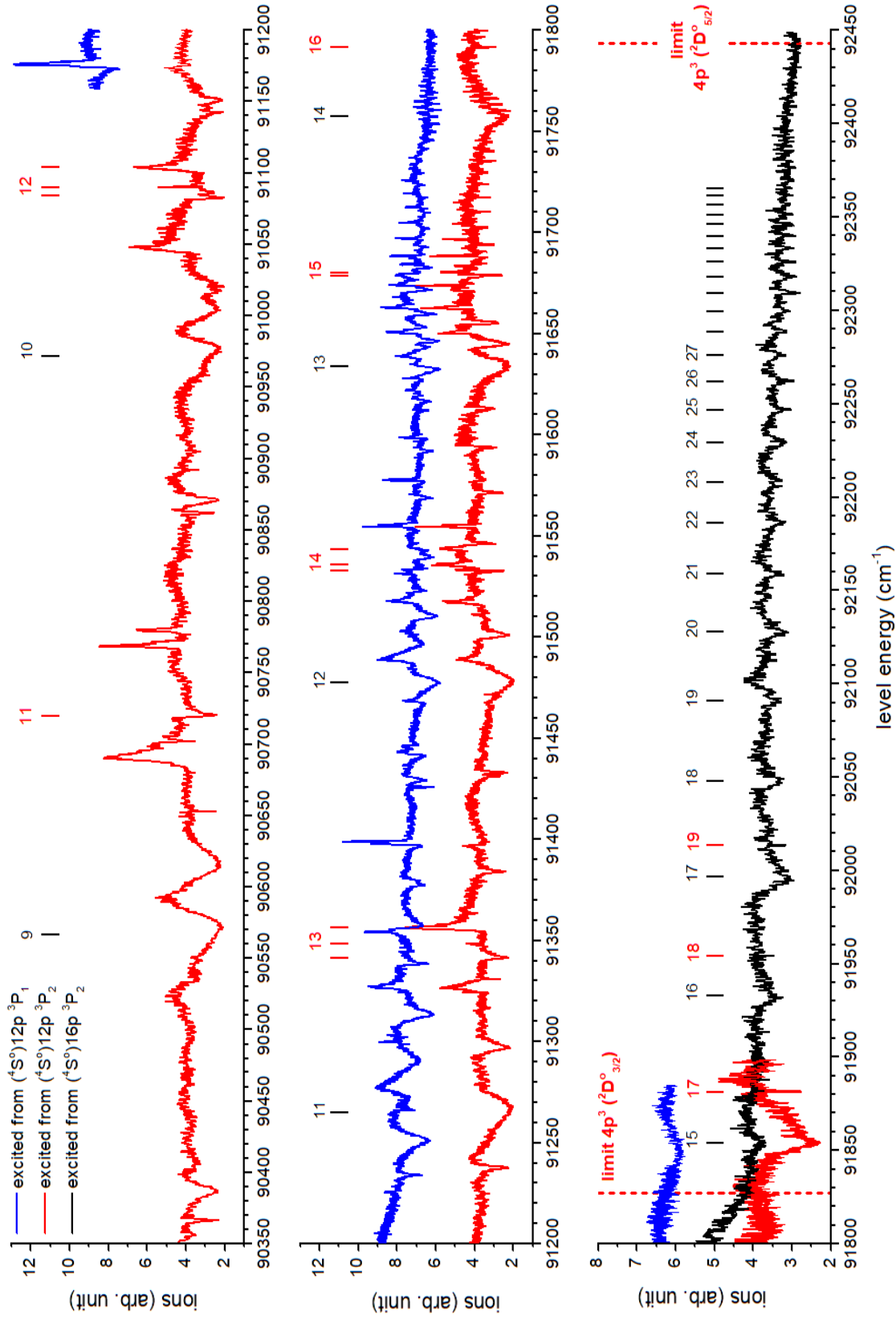

**Fig. 4.** Photoionization spectra of AI state series converging to the limit of core state $4s^2 4p^3 (^2D^o_{3/2})$ and $4s^2 4p^3 (^2D^o_{5/2})$ via scheme A1, A2 and A3. The Rydberg series converging to limit $4s^2 4p^3 (^2D^o_{5/2})$ are marked with "|": black color for $nd\ ^3D$ series and red color for $ns\ ^3D$.

As discussed in our previous work of Te [20], the analysis of AI states is challenging due to varying AI resonance profiles and width dependent on the interaction with underlying continuum. If considering two interacting channels, the continuum and the AI series of interest, the resonance can be described with a Fano profile:

$$I(\epsilon) = I_{res} \frac{(\epsilon+q)^2}{\epsilon^2+1} + I_{cont} \quad \text{with} \quad \epsilon = \frac{E - E_{res}}{\Gamma/2}, \tag{1}$$

Here $E_{res}$ is the energy of AI state; $\epsilon$ is the reduced energy determined by $E_{res}$ and the resonance width $\Gamma$; $I_{res}$ is the resonance intensity; $I_{cont}$ is the background signal caused by non-resonant ionization processes, e.g. the direct excitation from the intermediate level to the continuum; $q$ is the Fano parameter characterizing the resonance shape and asymmetry.

In this work, the photoionization spectra were obtained by exciting Se from the intermediate states $(^4S^o)12p$ or $16p$. Sharing the same core state, the excitation rate to the continuum configuration $(^4S^o)\varepsilon s$ and $\varepsilon d$ will be favored and comparable to the excitation rate to the embedded AI states. Depending on the competition of these two excitation paths, the profile of an AI series can change between Lorentzian, asymmetric, and window profiles. This makes fitting and grouping of the AI series challenging. Excitation from an intermediate state with a $(^2D^o)$ core would be desired, similar to our previous work on Te. However this scheme was infeasible in the present work limited by the wavelength range of Ti:Sa laser.

When analyzing the AI structure of chalcogens, the systematic similarities and trends with respect to the halogens and the noble gases can be helpful, since they all have $p$ orbitals as a valence shell [24]. Excited from $(^4S^o)12p$ or $16p$ intermediate states, the anticipated odd-parity AI Rydberg series will be the $4p^3(^2D^o_{5/2})ns$ and $nd$ series. For noble gases, halogens and chalcogens, except the light elements (O, F and Ne), there is a characteristic broad $nd$ AI series and a narrow $ns$ AI series converging to the first/second excited core state. For chalcogens and using $LS$ designations, the broad $nd$ series is designated as $nd\ ^3D$ and the narrow $ns$ series is assigned as $ns\ ^3D$. There are two more $nd$ series with narrow width: $nd\ ^3P$ and $nd\ ^3S$. The underlying continuum configurations are $(^4S^o)\varepsilon s\ ^3S$ and $\varepsilon d\ ^3D$. To autoionize, AI states $(^2D^o)ns\ ^3D$ and $(^2D^o)nd\ ^3S$ have to change their orbital angular momentum by two units, which significantly decreases the respective autoionization rate. Among all the narrow series, $nd\ ^3P$ has the smallest width since it is "forbidden" to autoionize by strict $LS$-coupling, and observed nonetheless by deriving strength only from spin-orbit interactions.

### 3.2.1 AI Rydberg ns and nd series converging to $4p^3\ (^2D^o_{5/2})$

The AI states converging to $4p^3\ (^2D^o_{5/2})$ were first analyzed due to the simplicity of the spectra in the energy range 91800-92450 cm$^{-1}$. Only one broad and one narrow resonance series were observed. The broad resonances show a window profile constant with a typical $(n^*)^{-3}$ width change. The narrow one may have a natural width smaller than that of the GHz-linewidth Ti:Sa lasers, and therefore shows a constant width (see Fig. 4). As discussed above, the broad resonances shall be designated as $nd\ ^3D$. For the member of $4s^24p^3(^2D_{5/2})nd\ ^3D$ with int[$n^*$]<14, the energy of the states are lower than the limit of $(^2D_{3/2})$, therefore they are immersed in a number of Rydberg series converging to $(^2D_{3/2})$. In this energy range significant perturbations were observed, which will be discussed in the following section. To simplify the fitting process, a Lorentzian/Gaussian profile was used for symmetric resonances depending on the resonance width; and the Fano profile was used for obviously asymmetric lines. The extracted level energies are presented in Tab. 3 and in Fig. 5 as a Lu-Fano plot. In the plot, $nd\ ^3D$ series shows a constant quantum defect mod$_1[\delta]$~0.35, which agrees with Cantú's result [14] on the same configuration at low-$n$ range. The deviation of their $\delta$ at high-$n$ range was attributed to wavelength inaccuracies. Gibson et al. [15] also measured this series but observed different results than Cantú et al. with quantum defect mod$_1[\delta]$~0.18. Both Cantú and Gibson's results are plotted in Fig. 5 for comparison. Referring to previous literature [14, 15,28], the principal number of the series $nd\ ^3D$ is assigned as $n$=int[$n^*$]+2.

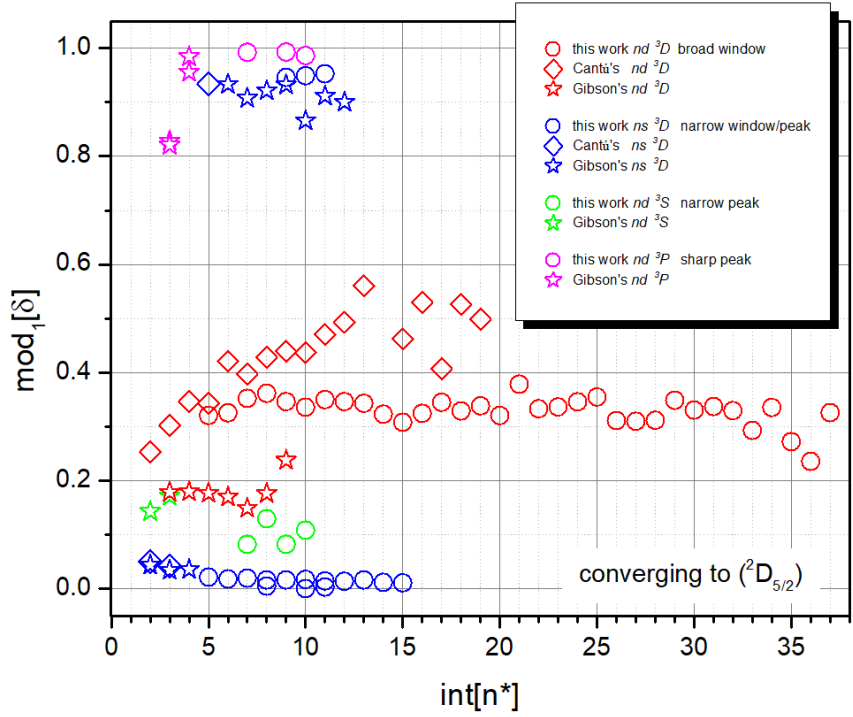

**Fig. 5.** Lu-Fano plot for the observed odd-parity Rydberg series converging to $4s^24p^3(^2D_{5/2})$. Previously reported level energies of the same series by Cantú et al. [14] and Gibson et al [15] are included for comparison. Gibson's assignments of $(^2D_{5/2})ns\ ^3D$ and $(^2D_{5/2})nd\ ^3S$ presented here has been corrected according to the later theoretical calculations [1].

With near constant quantum defect the $4s^24p^3(^2D_{5/2})nd\ ^3D$ series can be used to extract the energy of its converging limit as IP+13784.4 cm$^{-1}$. The extracted IP values will not be as precise as the one determined with the Rydberg series below the IP, limited by the precision of the ionic state energy 13784.4 [29] cm$^{-1}$ (the uncertainty was not explicitly claimed in [29], but should be ≥0.1 cm$^{-1}$) and the intermediate state energy 78060.03(15) cm$^{-1}$ measured in this work. Nevertheless it is still a good check on the accuracy. From the AI Rydberg series $4s^24p^3(^2D_{5/2})nd\ ^3D$ the IP is determined to be 78658.29(6)$_{stat}$(≥25)$_{sys}$ cm$^{-1}$, in agreement with the value extracted from Rydberg series. For the fitting, only the members above the $4s^24p^3(^2D_{3/2})$ limit were used, and *the* quantum defect $\delta$ was chosen as a constant $\delta_0$ without higher order terms.

The narrow resonance with mod$_1[\delta]$~0.01 cannot be assigned readily. It can be either $(^2D)ns\ ^3D$ or $(^2D)nd\ ^3S$. This goes back to the historical controversy about the assignment of the AI series $(^2D)ns\ ^3D$ and $(^2D)nd\ ^3S$ of S and Se [1]. The theoretical calculation preferred the assignment of $(^2D)ns\ ^3D$ at higher energy side of $(^2D)nd\ ^3S$ [1,26], which also agrees with our recent Te result of the corresponding AI series [20]. After revising Gibson's data according to the theoretical suggestion, both Gibson's and Cantú's data in Fig. 5 suggest these observed narrow resonances to be $(^2D_{5/2})ns\ ^3D$.

At the energy range lower than the limit $4p^3(^2D_{3/2})$, e.g., int[n*]<13 in Fig. 5, the low-n Rydberg members were observed using scheme A2. In this range, the fine structures of the $ns\ ^3D$ series (J=1,2,3) were observed: two window resonances with mod$_1[\delta]$~0.01 and one peak resonance with mod$_1[\delta]$~0.95. In addition, two more series were observed: one with narrow resonances and the other one with super sharp resonances. Following the discussion at the beginning of Sect. 3.2., the narrow resonances with mod$_1[\delta]$~0.10 are assigned as $nd\ ^3S$ and the super sharp resonances with mod$_1[\delta]$~0.99 are assigned as $nd\ ^3P$. All the AI Rydberg state energies are presented in Tab. 3.

**Table 3.** Series $4s^2 4p^3(^2D_{5/2})nd\ ^3D$, $ns\ ^3D$, $nd\ ^3S$ and $nd\ ^3P$ converging to the second excited state of Se II at 92442.6 cm$^{-1}$.

| n | $nd\ ^3D_{1,2,3}$ | | n | $ns\ ^3D_{1,2,3}$ | | | | | |
|---|---|---|---|---|---|---|---|---|---|
| | $E_n$ (cm$^{-1}$) | δ | | $E_n$ (cm$^{-1}$) | δ | $E_n$ (cm$^{-1}$) | δ | $E_n$ (cm$^{-1}$) | δ |
| 7 | 89040.2 | 1.32 | 9 | 89372.0 | 3.02 | | | | |
| 8 | 89979.1 | 1.33 | 10 | 90191.1 | 3.02 | | | | |
| 9 | 90566.2 | 1.35 | 11 | 90719.3 | 3.02 | | | | |
| 10 | 90971.9 | 1.36 | 12 | 91082.7 | 3.02 | 91086.1 | 3.01 | 91104.0 | 2.95 |
| 11 | 91265.0 | 1.35 | 13 | 91341.6 | 3.02 | | | 91356.3 | 2.95 |
| 12 | 91477.6 | 1.34 | 14 | 91532.7 | 3.02 | 91535.5 | 3.00 | 91543.4 | 2.95 |
| 13 | 91634.0 | 1.35 | 15 | 91678.6 | 3.01 | 91680.0 | 3.00 | | |
| 14 | 91757.2 | 1.35 | 16 | 91791.8 | 3.01 | | | | |
| 15 | 91854.2 | 1.34 | 17 | 91881.3 | 3.02 | | | | |
| 16 | 91933.1 | 1.32 | 18 | 91954.0 | 3.01 | | | | |
| 17 | 91996.9 | 1.31 | 19 | 92013.3 | 3.01 | | | | |
| 18 | 92047.9 | 1.32 | | | | | | | |
| 19 | 92090.5 | 1.35 | | | | | | | |
| 20 | 92127.8 | 1.33 | | | | | | | |
| 21 | 92158.7 | 1.34 | | | | | | | |
| 22 | 92185.9 | 1.32 | | | | | | | |
| 23 | 92207.8 | 1.38 | n | | | $nd\ ^3P_{1,2}$ | | | |
| 24 | 92229.0 | 1.33 | | | | $E_n$ (cm$^{-1}$) | δ | | |
| 25 | 92246.6 | 1.34 | 8 | | | 90207.9 | 0.99 | | |
| 26 | 92262.0 | 1.35 | 10 | | | 91090.0 | 0.99 | | |
| 27 | 92275.7 | 1.35 | 11 | | | 91348.2 | 0.99 | | |
| 28 | 92288.5 | 1.31 | | | | | | | |
| 29 | 92299.4 | 1.31 | | | | | | | |
| 30 | 92309.2 | 1.31 | | | | | | | |
| 31 | 92317.7 | 1.35 | | | | | | | |
| 32 | 92325.9 | 1.33 | | | | | | | |
| 33 | 92333.1 | 1.34 | n | | | $nd\ ^3S_1$ | | | |
| 34 | 92339.7 | 1.33 | | | | $E_n$ (cm$^{-1}$) | δ | | |
| 35 | 92346.0 | 1.29 | 9 | | | 90692.0 | 1.08 | | |
| 36 | 92351.2 | 1.34 | 10 | | | 91047.8 | 1.13 | | |
| 37 | 92356.6 | 1.27 | 11 | | | 91326.8 | 1.08 | | |
| 38 | 92361.4 | 1.24 | 12 | | | 91517.5 | 1.11 | | |
| 39 | 92365.2 | 1.33 | | | | | | | |

### 3.2.2 AI Rydberg series converging to $4p^3\ (^2D^o_{3/2})$

The photoionization spectra in the energy range below 91800 cm$^{-1}$ are more complicated due to the overlap of multiple AI series converging to the neighboring limits $4p^3\ (^2D^o_{3/2})$ and $4p^3\ (^2D^o_{5/2})$. Moreover, the complexity is prominently increased by strong perturbations in the range of 90400-91700 cm$^{-1}$, which largely deviate the quantum defects of some Rydberg series converging to the limit $(^2D_{3/2})$. These perturbations also cause resonance-shape change and further raise the difficulty in grouping the series. Fortunately, in the vicinity of the limit $4p^3\ (^2D^o_{3/2})$, obvious regularity is observed for two groups of resonances: one with peak profile and the other one with window profile. Since no obvious asymmetries were observed, these resonances could be fitted using Lorentzian/Gaussian profiles. The extracted state energies are presented in both Lu-Fano plot (Fig. 6) and Tab. 4.

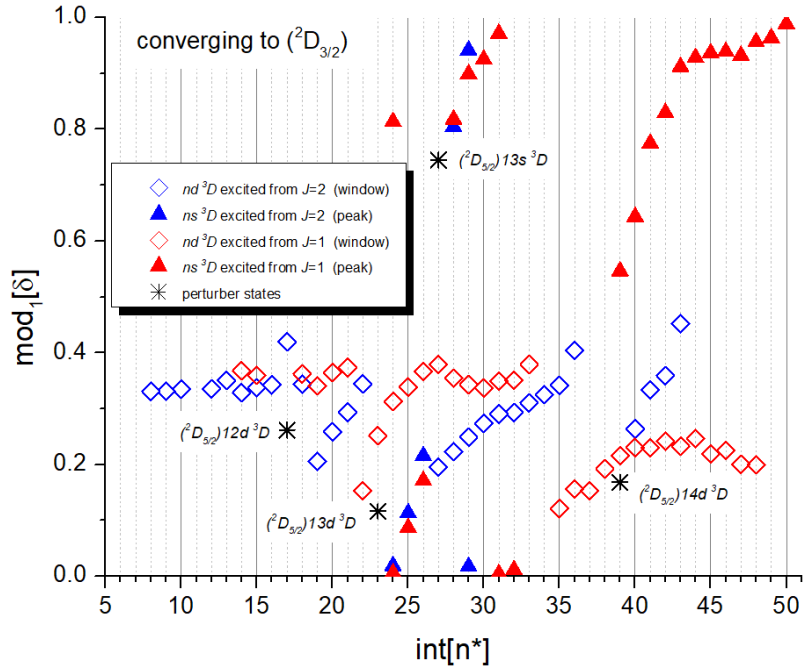

**Fig. 6.** Lu-Fano plot for the observed odd-parity Rydberg series converging to $4s^24p^3(^2D_{3/2})$. The level energies were measured using the photoionization spectra excited from the intermediate state $4s^24p^3(^4S^o)12p\ ^3P$ ($J=1,2$). The perturber states $(^2D_{5/2})nd\ ^3D$ ($n=12,13,14$) and $(^2D_{5/2})15s\ ^3D$ marked as "✳" were observed in the spectrum from the $J = 2$ intermediate state and measured in Sect. 3.2.1.

Similar to the assignment discussed in Sect.3.2.1, the window resonance series should be $(^2D_{3/2})nd\ ^3D$ with $mod_1[\delta]\sim 0.35$ and the peak resonance series should be $(^2D_{3/2})ns\ ^3D$ with $mod_1[\delta]\sim 0.00$. Similar peak and window resonance series are observed in both spectra excited from $(^4S^o)12p\ ^3P\ J=1$ and from $J=2$ as expected, but at non-negligibly different energies which cannot be attributed to the choice of fitting profile or the systematic energy shift of the intermediate state. Hence, the $nd$ and $ns$ series observed via different intermediate states are likely the fine-structure states of $(^2D_{3/2})nd\ ^3D$ and $(^2D_{3/2})ns\ ^3D$ with different $J$. As noted in Fig. 6, the two $nd$ series excited from the $J=2$ and $J=1$ intermediate states, respectively, are substantially perturbed at different energies. The $nd$ series excited from $J=2$ state (blue "◇" in Fig. 6) could be perturbed by the $(^2D_{5/2})nd\ ^3D$ ($n=12,13,14$) (marked by "✳" in Fig. 6) as they are located right at where the $\delta$ of the $nd$ series deviates. To perturb a Rydberg series, the perturber state must have the same $J$ as the series. Therefore, $(^2D_{5/2})nd\ ^3D$ ($n=12,13,14$) should have the same $J$ as the blue "◇" $nd$ series. On the other hand, for the $nd$ series excited from the $J=1$ intermediate state (red "◇" in Fig. 6) significant $\delta$ deviations happen around $int[n^*]=21$ and 34 instead. The perturber states could not be identified limited by the spectrum complexity. Better understanding of these observations requires more advanced theoretical calculations like multichannel quantum defect theory (MQDT) [30]. Due to a multitude of channels involved, relativistic multichannel theory (RMCT) [31] may have to be employed to calculate the MQDT parameters directly.

**Table 4.** Series $4s^24p^3(^2D_{3/2})nd\ ^3D$ and $ns\ ^3D$ converging to the limit of 91826.6 cm$^{-1}$. The level energies of this work were measured via scheme A1 and A2.

| | nd $^3D$ | | | | ns $^3D$ | | | |
|---|---|---|---|---|---|---|---|---|
| n | excited from J=1 | | excited from J=2 | | excited from J=1 | | excited from J=2 | |
| | $E_n$ (cm$^{-1}$) | δ | $E_n$ (cm$^{-1}$) | δ | $E_n$ (cm$^{-1}$) | δ | $E_n$ (cm$^{-1}$) | δ |
| 10 | | | 90366.2 | 1.33 | | | | |
| 11 | | | 90652.5 | 1.33 | | | | |
| 12 | | | 90861.6 | 1.34 | | | | |
| 13 | | | | | | | | |
| 14 | | | 91142.1 | 1.34 | | | | |
| 15 | | | 91237.4 | 1.35 | | | | |
| 16 | 91313.8 | 1.37 | 91316.5 | 1.33 | | | | |
| 17 | 91377.7 | 1.36 | 91379.0 | 1.34 | | | | |
| 18 | | | 91430.9 | 1.34 | | | | |
| 19 | | | 91471.3 | 1.42 | | | | |
| 20 | 91510.4 | 1.36 | 91511.1 | 1.34 | | | | |
| 21 | 91542.4 | 1.34 | 91546.3 | 1.21 | | | | |
| 22 | 91568.6 | 1.36 | 91571.3 | 1.26 | | | | |
| 23 | 91591.7 | 1.37 | 91593.4 | 1.29 | | | | |
| 24 | 91616.1 | 1.15 | 91612.6 | 1.34 | | | | |
| 25 | 91631.8 | 1.25 | | | | | | |
| 26 | 91646.3 | 1.31 | | | | | | |
| 27 | 91659.7 | 1.34 | | | 91638.8 | 2.81 | | |
| 28 | 91671.7 | 1.37 | | | 91650.7 | 3.01 | 91650.5 | 3.02 |
| 29 | 91682.5 | 1.38 | 91684.4 | 1.20 | 91662.9 | 3.09 | 91662.6 | 3.11 |
| 30 | 91692.6 | 1.35 | 91693.8 | 1.22 | 91673.9 | 3.17 | 91673.4 | 3.22 |
| 31 | 91701.6 | 1.34 | 91702.4 | 1.25 | 91688.2 | 2.82 | 91688.3 | 2.80 |
| 32 | 91709.6 | 1.34 | 91710.1 | 1.27 | 91696.8 | 2.90 | 91696.4 | 2.94 |
| 33 | 91716.8 | 1.35 | 91717.2 | 1.29 | 91705.0 | 2.93 | 91704.3 | 3.02 |
| 34 | 91723.4 | 1.35 | 91723.8 | 1.29 | 91712.4 | 2.97 | | |
| 35 | 91729.3 | 1.38 | 91729.7 | 1.31 | 91719.2 | 3.00 | | |
| 36 | | | 91735.1 | 1.33 | 91725.5 | 3.01 | | |
| 37 | 91741.1 | 1.12 | 91740.0 | 1.34 | | | | |
| 38 | 91745.5 | 1.16 | 91744.4 | 1.40 | | | | |
| 39 | 91749.7 | 1.15 | | | | | | |
| 40 | 91753.5 | 1.19 | | | | | | |
| 41 | 91757.0 | 1.22 | | | | | | |
| 42 | 91760.3 | 1.23 | 91760.2 | 1.26 | 91755.9 | 2.55 | | |
| 43 | 91763.5 | 1.23 | 91763.1 | 1.33 | 91759.0 | 2.64 | | |
| 44 | 91766.3 | 1.24 | 91766.0 | 1.36 | 91761.8 | 2.77 | | |
| 45 | 91769.1 | 1.23 | | | 91764.6 | 2.83 | | |
| 46 | 91771.6 | 1.25 | | | 91767.2 | 2.91 | | |
| 47 | 91774.0 | 1.22 | | | 91769.9 | 2.93 | | |
| 48 | 91776.2 | 1.23 | | | 91772.3 | 2.94 | | |

| | | | | | |
|---|---|---|---|---|---|
| 49 | 91778.3 | 1.20 | | 91774.6 | 2.94 |
| 50 | 91780.3 | 1.20 | | 91776.8 | 2.93 |
| 51 | | | | 91778.8 | 2.96 |
| 52 | | | | 91780.7 | 2.96 |
| 53 | | | | 91782.5 | 2.99 |

## 4. Summary

Using a multitude of excitation schemes, photoionization spectra of Se have been obtained. The Rydberg and AI Rydberg states converging to different limits $4s^24p^3(^4S_{3/2})$ (the IP), $4s^24p^3(^2D_{3/2})$ and $4s^24p^3(^2D_{5/2})$ were measured and assigned. From measured Rydberg series $4s^24p^3(^4S)np\ ^3P_2$, the IP was determined to be 76658.15(2)$_{stat}$(4)$_{sys}$ cm$^{-1}$, which is different from the currently accepted value of 78658.35(12) cm$^{-1}$ [2, 12], but agrees with the value of 78658.12(10) cm$^{-1}$ measured by Morillon and Vergès [10]. This resolves the historical discrepancy and improves the precision of the Se ionization potential.

**Acknowledgements**

The experimental work has been funded by TRIUMF which receives federal funding via a contribution agreement with the National Research Council of Canada and through a Natural Sciences and Engineering Research Council of Canada (NSERC) Discovery Grant (SAP-IN-2017-00039). Y. Liu acknowledges support from the U.S. Department of Energy, Office of Science, and Office of Nuclear Physics under contract number DE-AC05-00OR22725. M. Mostamand acknowledges funding through the University of Manitoba graduate fellowship.